\documentclass[iop]{emulateapj}
\shorttitle{Models for KSN 2015K}
\shortauthors{Tolstov et al.}

\usepackage{color}
\usepackage{enumerate}
\usepackage{amsmath}
\usepackage{textcomp}
\usepackage{longtable}
\usepackage{ulem}

\usepackage{scalefnt}

\begin{document}

%\title{Models for fast-evolving supernova KSN 2015K: \\ light curves of explosions of Super-AGB progentors}
\title{Light curve models for fast-evolving supernova KSN 2015K: \\ explosions of Super-AGB progentors}

\title{Light curve models for fast-evolving supernova KSN 2015K: \\ explosions in circumstellar matter of Super-AGB progentors}

\title{Light curve modeling of fast-evolving supernova KSN 2015K: \\ explosion in circumstellar matter of Super-AGB progentor}

\author{Alexey Tolstov\altaffilmark{1}, Ken'ichi Nomoto\altaffilmark{2}, Elena Sorokina\altaffilmark{3,4},  
Sergei Blinnikov\altaffilmark{4,2,5}, Nozomu Tominaga\altaffilmark{6}, Yoshiaki Taniguchi\altaffilmark{1}
}

\affil{\altaffilmark{1}
The Open University of Japan, 2-11, Wakaba, Mihama-ku, Chiba, Chiba 261-8586, Japan}

\affil{\altaffilmark{2} Kavli Institute for the Physics and Mathematics of the Universe (WPI), The
University of Tokyo Institutes for Advanced Study, The University of Tokyo, 5-1-5 Kashiwanoha, Kashiwa, Chiba 277-8583, Japan} 

\affil{\altaffilmark{3} Sternberg Astronomical Institute, M.V.Lomonosov Moscow State University, 119234 Moscow, Russia}

\affil{\altaffilmark{4} NRC "Kurchatov Institute" - ITEP, 117218 Moscow, Russia} 

\affil{\altaffilmark{5} Space Research Institute (IKI), 117997 Moscow, Russia} 

\affil{\altaffilmark{6} Department of Physics, Faculty of Science and Engineering, Konan University, 8-9-1 Okamoto, Kobe, Hyogo 658-8501, Japan}

\email{$^{*}$ E-mail: alexey.tolstov@ouj.ac.jp}

\submitted{Accepted for publication in the Astrophysical Journal Letters}
\journalinfo{Accepted for publication in the Astrophysical Journal Letters}
\slugcomment{Accepted for publication in the Astrophysical Journal Letters on 06 Jun 2019}

\begin{abstract}
\noindent

Recent supernova searches revealed a number of fast-evolving luminous transients. We perform radiation hydrodynamic simulations  of light curves of several models of supernova from super--asymptotic giant branch (super-AGB, SAGB) stars with low mass envelopes ($M_{\rm env}$ $\sim$ $0.05-1.25$ $M_{\odot}$). 
The differences in the light curves found among the models are 
used to link the observed events to the most appropriate models.
In particular, we propose that KSN 2015K is an electron-capture supernova. We assume "optically" thick CSM around SAGB and the circumstellar interaction powers the peak luminosity of the light curve with a short rise time. The faint tail might be influenced by the spin-down luminosity of a newborn Crab-like pulsar.
 Our fits indicate an ejecta mass of $0.02-0.05$ $M_{\odot}$, a circumstellar medium (CSM) mass of $0.10-0.12$ $M_{\odot}$, a radius of the CSM photosphere $\sim$ 10$^{14}$ cm, a kinetic energy of $\sim$ 3 $\times$ 10$^{50}$ erg, a photospheric velocity $v_{\rm ph} \gtrsim 10,000$ km s$^{-1}$ and a pulsar total spin energy ($2.5-4$) $\times$ 10$^{49}$ erg.

\end{abstract}


%=============================

\section{INTRODUCTION}
\label{sec:intro}
\noindent

%%%%% Intro

Recently detected fast-evolving
luminous transients (FELTs) have peak luminosities comparable to Type Ia supernovae (SNe Ia) and the duration of their observations is about several tens of days. KSN 2015K is one of this kind of objects with even more extreme characteristics \citep{Rest2018}. KSN 2015K is in a star-forming spiral galaxy at redshift $z=0.090$. It has a very short rise time of 2.2 days and a time above half-maximum of only 6.8 days, while the typical values for observed FELTs are several times larger. 
The absolute magnitude at V-band maximum of KSN 2015K is $-18.8$ mags. After maximum light, it shows a decline
followed by a plateau and, finally, a power-law decay. The color at peak brightness is $r-i=-0.15\pm0.05$, and
$\sim$ 8 days after peak brightness its color remains quite blue at $g-r$ = $-0.17\pm0.20$.

%%%%%%%%%%%%%%%%%%%%%%%%%%%%%%%%%%%%%%%

%%% Possible models

At the moment, there is no universally accepted model for FELTs. The possible scenarios include ejected radioactive $^{56}$Ni, pulsar pumping, black hole accretion, or gamma-ray burst afterglow. All these scenarios being applied to KSN 2015K are hardly possible mostly due to the combination of the observed short rise-time and luminous peak. The scenarios are discussed in details in \citet{Rest2018}, including another possibility of KSN 2015K being powered by the interaction with CSM. They found that the interaction scenario could be most promising in explaining KSN 2015K by means of a numerical radiation hydrodynamic simulation with a grey flux-limited non-equilibrium diffusion approximation.

\citet{Rest2018} showed that KSN 2015K is most consistent with a shock breakout into a dense circumstellar
shell. 
% Our approach
One of the most striking examples to the low mass supernova explosion with a high mass-loss rate is an electron capture supernova (ECSN) \citep{Nomoto1984,Nomoto1987,Tominaga2013,Moriya2014}. Its progenitors has an O+Ne+Mg core, a red-supergiant envelope of a SAGB star, and a dense circumstellar shell surrounding the star which has been lost by various mechanisms such as dynamical pulsations, strong wind due to carbon dust, magnetic hearing, etc. While the mechanism(s) of mass loss 
from ECSN progenitors remains
uncertain, first-principle numerical simulations of the
O+Ne+Mg core collapse has predicted a small explosion energy ($\sim 10^{50}$ erg) and small production of $^{56}$Ni ($\sim 10^{-3}$ $M_{\odot}$) \citep{Kitaura2006,Burrows2007,Janka2008,Hoffman2008,Wanajo2009}.

In this paper, using multicolor radiation hydrodynamic
simulations, we perform the comparison of the  
light curves of KSN 2015K with several ECSN models interacting with CSM. In contracts to previously published simulations, our approach uses multigroup radiative transfer and more realistic SAGB progenitor models. The main purpose is to find
out whether ECSN models are promising to explain the extreme characteristics of KSN 2015K.

\section{Models}
\label{sec:models}
%\subsection{Methods}
\noindent

Similar to the approach described in \citet{Tominaga2013}
we take an O+Ne+Mg core model with 1.377 $M_{\odot}$ at a
presupernova stage from \citet{Nomoto1982} and \citet{Nomoto1984,Nomoto1987}. The model is a core of a star with $M_{\rm MS}$ =
$8.8M_{\odot}$ which is calculated from an He star with $2.2M_{\odot}$.
A mass range of stars with the O+Ne+Mg core is $M_{\rm MS} \sim 7--9.5M_{\odot}$, but a progenitor of the ECSN should possess an SAGB
envelope \citep[see][for a review]{Langer2012}, of which mass
and abundance are influenced by $M_{\rm MS}$, mass loss, and third
dredge-up associated with thermal pulses. However, the mass loss mechanism and its rate are highly
uncertain and no calculation of the full thermal pulses through the
presupernova has been available.
Our choice of $8.8M_{\odot}$ SAGB model of the core is rather arbitrary (it is the boundary mass between the ECSN and the ONeMg white dwarf formation).  A star in the 8-10 M$_{\odot}$ range forms a strongly degenerate ONeMg core with a very thin (in mass) He shell and very extended H-rich envelope on the SAGB.  Such an extended H-rich envelope does not affect the core structure, so that the evolution of the ONeMg core converges to almost the same route, especially, near the Chandrasekhar mass, irrespective of the initial mass.  The envelope structure depends on the ONeMg core mass (through the core mass - luminosity relation) and the envelope mass, but does not depend on the initial mass. These properties are exactly the same as in AGB stars developing the degenerate CO core, which does not depend on the initial mass of the 3--8 M$_{\odot}$ range.

\begin{figure}
\includegraphics[width=80mm]{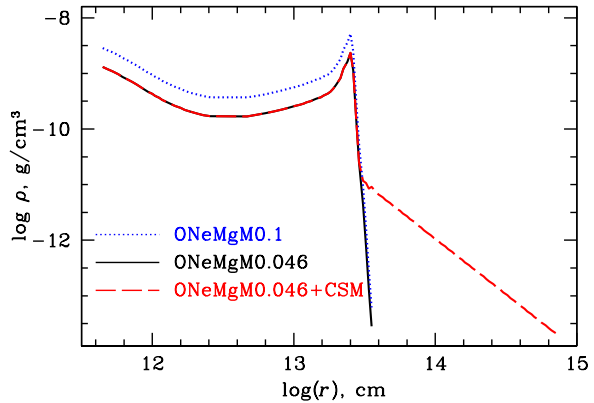}
\caption{ 
Initial density profiles for low-mass models. The dotted and solid lines show the models with the total mass of the ejecta $M_{\rm env}$ = 0.1 $M_{\odot}$ and $M_{\rm env}$ = 0.046 $M_{\odot}$, correspondingly. The dashed line shows 0.046 $M_{\odot}$ ejecta mass model with windy-like CSM and the total mass 0.1 $M_{\odot}$. 
\\}
\label{f1}
\end{figure}

In this paper we use two models
adopting such small envelope masses as $M_{\rm env}$ =
$0.046M_{\odot}$ and $M_{\rm env}$ = $1.25M_{\odot}$. In these models
we construct hydrostatic and thermal equilibrium envelopes. To cover
the mass range between $M_{\rm env}$ = $0.046M_{\odot}$ and
$1.25M_{\odot}$ we scale the mass of initial models by density variation (Figure \ref{f1}). 

Such small $M_{\rm env}$ implies 6.2 - 7.4 $M_\odot$ has been lost from the
8.8 $M_\odot$ progenitor, and the presupernova star is surrounded 
by a rather dense circumstellar medium (CSM).  
For all of our models, the outer radius $R_{\rm CSM}$ of the CSM is 
 $\sim 10^3 - 10^5 R_{\odot}$, or $\sim 10^{14}-10^{16}$ cm. The mass
loss process from SAGB stars is uncertain. We use a power-law density
distribution $\rho \propto r^{-p}$ for the CSM, which simulates the
wind that surrounds the exploding star. For a steady wind, $p = 2$, but
in the very last stages of the evolution of a presupernova star the
wind may not be steady (we varied $p$ in the range from
0 to 3.5). Assuming small velocity of the wind $v_{\rm
  {wind}} \sim 10$ km s$^{-1}$, the mass-loss rates in our models are
around $\sim10^{-3}-10^{-1}$ $M_{\odot}$ yr$^{-1}$ corresponding to
high mass-loss rate of Super-AGB stars.
Our SAGB progenitor lies just above the boundary mass between 
the ECSN and the ONeMg white dwarf formation \citep{Nomoto1984}.   

Chemical elements in the wind are supposed to be
distributed uniformly and they have the same abundances as the external layers of ejecta. We use hydrogen abundances $X_{\rm env}$(H) $\sim 0.7$ that is typical to SAGB star exteriors \citep{Jones2013}.
We also add some elements with higher atomic numbers
(usually 2\% of the total mass) with the abundances in solar
proportion. All models initially have T = 2.5 $\times$ 10$^3$ K in the wind. Higher temperatures produce an artificial flash of light
emitted by the huge CSM during its cooling \citep{BlinnikovSorokina2010}. Figure \ref{f1} demonstrates the density profile of the model with M$_{\rm env}$ = 0.046 $M_{\odot}$ and optional CSM. Here the outer layer of the envelope shows the density inversion which is formed in the super-adiabatic convective
layer of the SAGB star \citep{Nomoto1972}.

The explosion is initiated by a thermal bomb with the explosion energy
around values derived by the first-principle simulation ($E = 1.5 \times$ 10$^{50}$ erg; e.g., \citet{Kitaura2006}), producing a shock
wave that propagates outward. The subsequent evolution is followed by
a multi-group radiation hydrodynamic code STELLA
\citep{Blinnikov1998,Blinnikov2000,Blinnikov2006}.  The energy
deposition rate is $L_{\rm dep} = E_{\rm dep}/t_{\rm dep}$ 
during relatively short time $t_{\rm dep} \sim$ 0.1s.

We also investigate the contribution from the pulsar spin-down
luminosity that could be bright at birth. 
Since the pulsar spin-down luminosity can vary
in individual supernovae and the deposition efficiency of the energy released from the pulsar is unknown, we expediently adopt the initial spin-down luminosity of the Crab pulsar and assume
that the released energy is fully deposited at the bottom of
the ejecta (``full deposition'') 
%or deposited pursuant to the
%same one-group transport as gamma-rays from the radioactive decay
%(``one-group transport''). 
The energy deposition rate $L_{\rm dep}$ in
the presence of pulsar contribution is \citep{Kasen2010}
\begin{equation}
L_{\rm dep}=\frac{E_m/t_m}{(1+t/t_m)^2}.
\label{eqn1} 
\end{equation}
Here the total spin energy $E_m$ and spin-down timescale $t_m$ are 
connected with the pulsar spin period $P$ and its magnetic field $B$ as
\begin{equation}
\label{eqn2}
E_m \approx 2 \times 10^{52} P^{-2}_{\rm ms} {\rm ergs} , 
\end{equation}
\begin{equation}
\label{eqn3}
t_m \approx 5B^{-2}_{14} P^{2}_{\rm ms} {\rm days} ,
\end{equation}
where $P_{\rm ms}$ = $P/1$ ms and $B_{14}$ = $B/10^{14}$ G.
\\
\\
\\

%%%%%%%%%%%%%%%%%%%%%%

\begin{figure}
\includegraphics[width=80mm]{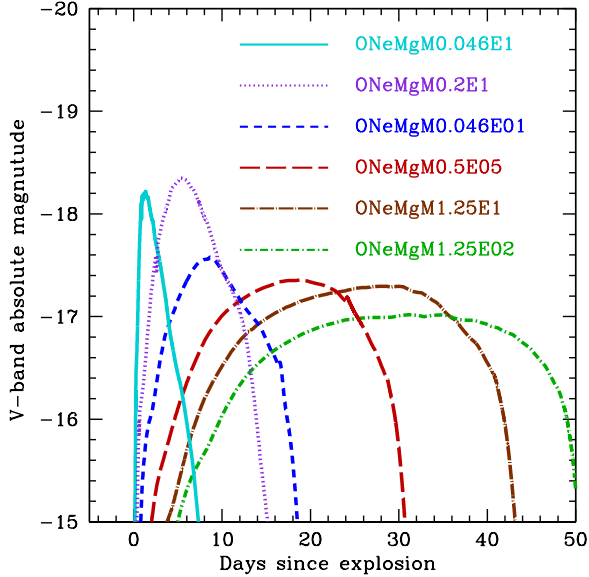}
\caption{
V-band light-curve simulations at low-mass ejecta range ($M \sim$ 0.05
-- 1.25 $M_{\odot}$). The names of the models include the ejecta mass
($M_{\odot}$), and released energy ($10^{51}$ erg).
\\}
\label{f2}
\end{figure}

\section{SIMULATIONS}
\label{sec:sims}
\noindent
\subsection{Light curves in the models with ejecta mass $M \sim$ 0.05 --- 1.25 $M_{\odot}$}

First of all, we checked whether the construction of low-mass hydrostatic and thermal equilibrium envelopes with mass $M \sim 0.05 -
1.25 M_{\odot}$ is suitable to explain FELTs light curves. In
Figure \ref{f2} a selected set of V-band light curves is
presented. The models with ejecta mass $M \sim 1 M_{\odot}$ have a
rising time $>10$ days and the duration of the plateau phase 
$\sim 10-30$ days. They can hardly be used to explain FELTs. The models
with the ejecta mass around $\sim 10^{-1} M_{\odot}$ are much more
promising: the rising time is only several days and the models fade from view in less than one month.

Among simulated light curves we have not found any good one to
reproduce the shape of the KSN 2015K light curve. All the light curves
have either larger rising time or smaller width of the peak. To
increase the width of the peak and make the rising time consistent with
observations, we have added a wind-like circumstellar medium to the
envelope (Figure \ref{f1}).

\subsection{CSM and $^{56}$Ni}

\begin{figure}
\includegraphics[width=80mm]{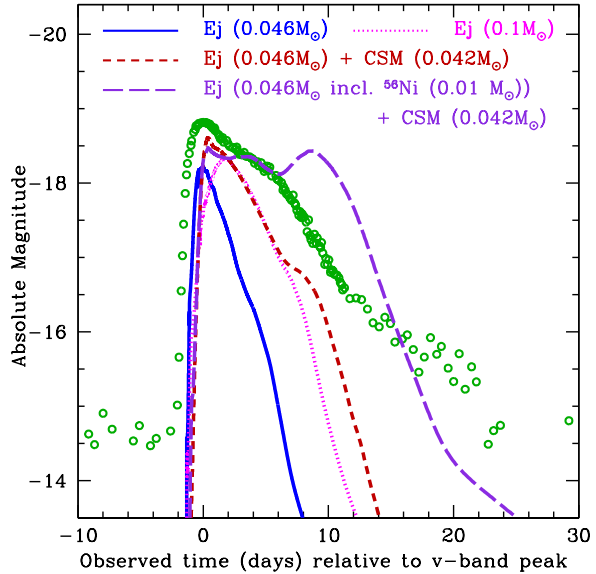}
\caption{K2 light-curve simulations for KSN 2015K in the models with CSM and $^{56}$Ni. In brackets masses of ejecta, CSM, and $^{56}$Ni are specified. Green dots denote KSN 2015K K2 light curve.
 \\}
\label{f3}
\end{figure}

In Figure \ref{f3} several models with CSM are presented. The presence
of CSM makes the light curve width broader owing to the increased mass and
extended propagation of the shock wave. It also allows to have short
rise time because the CSM is not so optically thick ($\tau\sim 0.1-1000$)
in comparison with the envelope ($\tau\sim 10^4-10^7$). 
Adding CSM helps to reproduce the shape of the light
curve of KSN 2015K and seems to be perspective for more detailed
modeling.

The models with CSM are good to reproduce the rising time and half-width
peak of KSN 2015K, but not the tail of the light curve. Interaction
models with CSM are usually declining too fast \citep[see
  also][]{Sorokina2016}.  It is possible that radioactive $^{56}$Ni
decay contributes to the luminosity at later times ($t > 10$ days). We
checked this by adding 0.01 $M_{\odot}$ of $^{56}$Ni in the inner
zones of ejecta, but the luminosity of the tail is fainter than 
observations. The largest amount of $^{56}$Ni is too
extreme. Moreover, the second peak appears in the light curve because of 
the $^{56}$Ni-decay heating of the ejecta.  Thus, the light curve tail can
hardly be explained by the radioactive $^{56}$Ni. It is more probable that
the tail is powered by continuing energy deposition from a central
remnant (a pulsar or black hole).  Another possibility that the tail
is formed by extended CSM with $R_{\rm CSM} > 10^{15}$ cm
\citep[see][]{Moriya2014}, but in this case the rising time becomes
too large to fit observations.  

\subsection{Best-fit models}

\begin{figure}
\includegraphics[width=80mm]{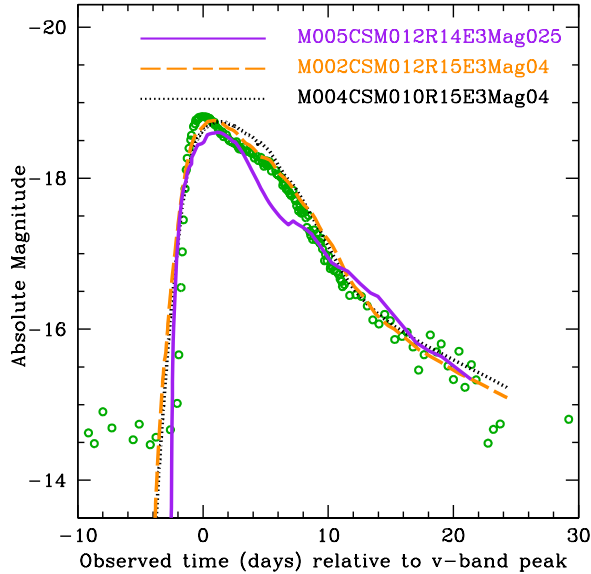}
\caption{Best-fit K2 light-curve simulations of KSN 2015K for 
  different models. The model names include the ejecta mass M
  ($0.02 - 0.05$ $M_{\odot}$), CSM mass ($0.10 - 0.12$ $M_{\odot}$), logarithm of the radius R of
  the CSM (cm), released energy E ($10^{50}$ erg), and pulsar total spin
  energy Emag ($10^{50}$ erg). Green dots denote KSN 2015K K2 light curve.
 \\}
\label{f4}
\end{figure}

Among $\sim$ 100 interaction models, we found several models whose 
rising time, peak luminosity and the decline rate are similar to the
observations of KSN 2015K (Figure \ref{f4}). These models have a rather small
ejecta mass $0.02-0.05$ $M_{\odot}$, a CSM mass of $0.10-0.12$
$M_{\odot}$, a radius of $\sim$ 10$^{14} - 10^{15}$ cm, a kinetic
energy of $\sim$ 3 $\times$ 10$^{50}$ erg, and energy supply ($2.5-4$)
$\times$ 10$^{49}$ erg from a central remnant forming the tail of the
light curve.  If the tail of the light curve is powered a pulsar, our
estimations shows that it is a Crab-like pulsar with the pulse period 
$P \sim 20$ ms and magnetic field $B \sim 2 \times 10^{12}$ G.

In the model with $R_{\rm CSM} \sim$ 10$^{14}$ cm, 
the shortest rising time in V-band is $\sim$ 3.2 days which
is longer than the observation (2.2 days), while the rising rate is
close to the observation.
The rising time in UV-bands is as short as 2 days, and only
0.1 days in X-rays. The too long rising time in V-band can be related
to the opacity or asymmetry effect and should be studied elsewhere.

\begin{figure}
\includegraphics[width=80mm]{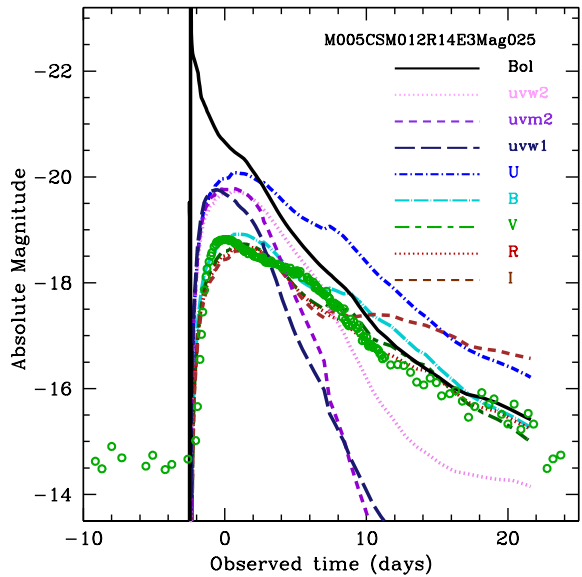}
\caption{Bolometric (black line) and multicolor light-curve
 simulations of KSN 2015K in the interaction model
 M005CSM012R14E3Mag025. Green dots denote the observed K2 light curve of KSN  2015K.            
 \\}
\label{f5}
\end{figure}

\subsection{Multicolor light curves}

We present the multicolor light curves for one of the best-fit model
(Figure \ref{f5}). The peak of the bolometric light curve is formed by
soft X-ray emissions ($\sim$ 100 eV). The UV- and U- bands are also brighter
than the optical bands by $\sim$ 1 mag. Bright U- and UV-
bands light curves at the peak are the consequence of high color
temperature $T \sim 10^4$ K. For KSN 2015K there is no data for those
bands, and more detailed future observations of similar objects can be
very useful to reveal the nature of FELTs.

For both models with the envelope and with CSM, the color at
peak brightness is rather blue ($g-r \sim -0.3$) and remains blue at $g-r \sim -0.2$
for $\sim 8$ days after peak brightness.  The blue color is also
observed in KSN 2015K.

For all the models the photospheric velocities are as high as $v_{\rm ph}
\gtrsim 10,000$ km s$^{-1}$. For models with CSM, $v_{\rm ph}$ remains high
at $+20$ days, while for models without CSM they become $v_{\rm ph} < 10,000$
km s$^{-1}$ already at $t > 10$ days. Thus, the relatively long
lasting evolution of the velocity can be an indicator of CSM.

We did not reveal any
significant difference in optical light curves by variation of
power-law in density distribution of optically thin CSM. In case of
steep density profile, more efficient acceleration of the shock wave
can be identified in X-rays and UV-band light curves.
The identification of the progenitor scenario
is an important issue, but it can be resolved only in more detailed
multi-band simulations and using all wavelengths observations from
early rise to late decline epoch.

\subsection{Smearing}

In one-dimensional simulations, the shell between the SN ejecta and the
dense CSM becomes thin and dense. But in reality and 3D simulations
such a shell is unstable. In the STELLA code, we take the
multidimensional effects into account by introducing a smearing term
$B_{\rm q}$ in the equation of motion to reduce the conversion efficiency
from kinetic energy to radiation energy \citep{Blinnikov1998,
  Moriya2013}.  In our simulations we used the standard value for
smearing term $B_{\rm q}=1$. We found that the optical light curve peak of low-mass ejecta models is
very sensitive to the value of smearing term, while smearing term has no significant effect on the late time light curves.  The smearing term produces the difference
in visual peak magnitude up to 2 mags. Bolometric peak is not affected
so much by smearing term giving the difference less than 0.5
mag. Thus, multi-dimensional simulations are needed to calibrate this
parameter in multicolor simulations.

\section{CONCLUSIONS AND DISCUSSION}
\label{sec:conclusion}

Using detailed radiation-hydrodynamic simulations of SAGB star explosions, 
we have found several best-fit models for the observed multicolor light curves of KSN 2015K.
Our best-fit models are consistent with a shock
breakout in a dense circumstellar shell and indicate the ejecta mass
of only a few times $\sim$ 10$^{-2} M_{\odot}$, CSM mass of
0.10--0.12 $M_{\odot}$, photospheric radius of $\sim$ 10$^{14}$ cm, and kinetic
energy of $ \sim$ 3 $\times$ 10$^{50}$ erg.

The tail luminosity of KSN 2015K declines by $\sim 0.5$ mag day$^{-1}$
for the model without the pulsar contribution. Interestingly, if the Crab-like
pulsar contributes to the light curve, the decline rate is lowered to
be in good agreement with observations. The magnetar in our model is required to explain the tail of light curve, but not the peak (in contrast to \citet{Rest2018} estimations). Crab-like pulsar can be a source of radiation at the late times. The observed decline rate can hardly be explained by
the $^{56}$Ni-decay only because it requires the mass of $^{56}$Ni comparable with
the mass of ejecta. Even a small amount of $^{56}$Ni ($\sim 0.01 M_{\odot}$) 
heats the low-mass ejecta and strongly affects on the shape of the
light curve.

%%% CSM model and short conclusion from Rest2018
Our best-fit parameter values are similar to those what proposed by
\citet{Rest2018}, where the progenitor is a compact (presumably
stripped envelope) star with a radius of $\sim 10^{11}$ cm surrounded
by a dense CSM that extended to radii of several times 10$^{14}$
cm. They modeled the supernova ejecta as simply a homologously expanding
broken power-law and the CSM as a constant-density shell. The
best-fit parameter values in \citet{Rest2018} reveal the ejecta mass
$M_{\rm ej}$ = 10 $M_{\odot}$, the outer ejecta velocity $V_{\rm ej}$ =
8500 km s$^{-1}$, mass of CSM $M_{\rm CSM}$ = 0.15 $M_{\odot}$, radius
of CSM $R_{\rm CSM}$ = 4 $\times$ 10$^{14}$ cm, and the shell width
$\Delta R_{\rm CSM}$ $\sim$ 0.25$R_{\rm CSM}$.
The probable evolutionary scenario of the \citet{Rest2018} model might be a stripped-envelope SN from a massive Wolf-Rayet (WR) star-like progenitor, 
although \citet{Rest2018} did not discuss this point.
If the supernova from a massive WR star produced $\sim 0.1 M_{\odot}$ $^{56}$Ni, 
its light curve should have a much broader peak than KSN 2015K.  
Therefore, the WR SN model must have produced negligible amount of $^{56}$Ni.

The CSM formed from the WR and SAGB progenitors must have some differences.
Because of the difference in the radius and core mass, thus in the
escape velocity, the expansion velocities of CSM are much higher
in the WR progenitor than the SAGB progenitor.
Thus the formation of the 0.15 $M_{\odot}$ CSM with $R_{\rm CSM}$ = 4 $\times$ 10$^{14}$ cm
having high velocities would take place in less than a year in dynamical fashion.

In contrast to the above model by \citet{Rest2018}, we do not assume
such a high ejecta mass as 10 $M_{\odot}$ and we use the more
realistic progenitor and explosion models constructed from the
evolutionary models of SAGB stars and the simulations of ECSNe.
The CSM of SAGB stars is the result of more like a slow wind,
although the mass loss rate must be as high as so-called super-wind.
As mentioned in \S \ref{sec:models}, our SAGB progenitor lies just above the boundary mass between 
the ECSN and the ONeMg white dwarf formation \citep{Nomoto1984}.

Due to many uncertainties, such as the mass loss mechanisms and the                            
mass loss rate, it is not so easy to use predicted rates of SNe for
discrimination between SAGB stars and massive WR stars. \citet{Mathews2014} gives an estimate for ONeMg SNe about 10\% of all CCSNe. Ibc SNe constitute about 10\% of all SNe \citep{Leaman2011}, but ultra-stripped SNe constitute only a small fraction of all SNe Ic \citep{Tauris2013}. FELTSs are observed in 4-7\% of all CCSNe \citep{Drout2014} and they seem to be not limited by one scenario.

We have modeled FELTs in one-dimensional simulations. The aspherical
effects are not taken into account, but could be important both for low-mass SNe and WR SNe.
Future multi-dimensional radiation-hydrodynamic calculations are
needed to investigate more accurately the effects of
asphericity. Multi-dimensional simulations can also clarify the
smearing effect that have a large impact on the peak luminosity of the
multicolor light curves.

FELTs are difficult to discover and follow up, but in recent years the
number of their detections is growing on such surveys as the Panoramic
Survey Telescope and Rapid Response System 1 (PS1), Palomar
Transient Factory (PTF). To find out the nature of these supernovae,
both more detailed numerical simulations and future follow-up
observations in soft X-rays, UV- and U-band wavelengths are highly in demand.
 
\acknowledgments

This research is supported by the World Premier International Research Center Initiative (WPI Initiative), MEXT, Japan, and JSPS KAKENHI Grant Numbers JP16K17658, JP26400222, JP16H02168, JP17K05382, JP16H02166. The work of ES (light curve calculation with extended line list) was supported by the Russian Scientific Foundation grant 16--12--10519. The work of SB on development of STELLA code is supported by the grant RSF 18--12--00522.

\bibliographystyle{apj}
\bibliography{bibfile}

\end{document}